\documentclass[twocolumn,prb,superscriptaddress]{revtex4}

\usepackage{picinpar}
\usepackage{array}
\usepackage{subfigure}
\usepackage{amsmath}
\usepackage{multirow}
\usepackage[dvips]{graphicx}
\usepackage{amssymb}

\begin{document}

\title{Electronic levels and electrical response of periodic molecular structures from plane-wave orbital-dependent calculations}

\author{Yanli Li}
\affiliation{Department of Physics, Xiamen University, Xiamen 361005, Republic of China}
\author{Ismaila Dabo}
\email{daboi@cermics.enpc.fr}
\affiliation{Universit\'e Paris-Est, CERMICS, Projet Micmac ENPC-INRIA, 
6-8 avenue Blaise Pascal, 77455 Marne-la-Vall\'ee Cedex 2, France}


\begin{abstract}
Plane-wave electronic-structure predictions based upon orbital-dependent density-functional theory (OD-DFT) approximations, such as hybrid density-functional methods and self-interaction density-functional corrections, are severely affected by computational inaccuracies in evaluating electron interactions in the plane-wave representation. These errors arise from divergence singularities in the plane-wave summation of electrostatic and exchange interaction contributions. Auxiliary-function corrections are reciprocal-space countercharge corrections that cancel plane-wave singularities through the addition of an auxiliary function to the point-charge electrostatic kernel that enters into the expression of interaction terms. At variance with real-space countercharge corrections that are employed in the context of density-functional theory (DFT), reciprocal-space corrections are computationally inexpensive, making them suited to more demanding OD-DFT calculations. Nevertheless, there exists much freedom in the choice of auxiliary functions and various definitions result in different levels of performance in eliminating plane-wave inaccuracies. In this work, we derive exact point-charge auxiliary functions for the description of molecular structures of arbitrary translational symmetry, including the yet unaddressed one-dimensional case. In addition, we provide a critical assessment of different reciprocal-space countercharge corrections and demonstrate the improved accuracy of point-charge auxiliary functions in predicting the electronic levels and electrical response of conjugated polymers from plane-wave OD-DFT calculations.
\end{abstract}

\maketitle

\section{Introduction}

Determining the electronic levels and electrical response of materials stands as one of the fundamental limitations of density-functional theory (DFT) approximations. \cite{PayneArias1992} To address this deficiency, orbital-dependent density-functional theory (OD-DFT) approximations represent promising alternatives. \cite{KummelKronik2008} At present, the most widely used OD-DFT methods are hybrid functional approximations that consist of admixing a fraction of Hartree-Fock (HF) exchange into DFT functionals. \cite{Becke1993} In hybrid functionals, orbital dependence manifests itself into the nonlocal Fock contribution to the admixed Hamiltonian that causes the effective potential to vary with the orbital state upon which it acts. Hybrid functional methods are now well established in the field of electronic-structure calculations and have been shown to improve upon local and semilocal DFT in predicting orbital properties. In parallel, self-interaction corrections to DFT approximations represent a second category of OD-DFT methods that has rapidly grown in recognition. In the self-interaction approach, the total energy of the system is expressed explicitly in terms of individual orbital densities, yielding orbital-dependent effective potentials. This orbital specificity is then exploited as an additional degree of freedom to correct unphysical errors inherent in the orbital-independent picture. Although progress has been gradual since the introduction of self-interaction corrections, \cite{Perdew1981} number of successful applications have appeared recently. \cite{KummelKronik2008}

Despite the predictive potential of OD-DFT methods, conventional plane-wave OD-DFT implementations suffer from important computational inaccuracies in comparison with predictions based upon atomic Slater functions, local Gaussian orbitals, semilocal wavelets, \cite{GenoveseDeutsch2006} specific discrete representations, \cite{LiuYarne2003} and real-space finite differences. \cite{NatanBenjamini2008} These inaccuracies, which are not intrinsic to the plane-wave representation, arise from divergence singularities in the reciprocal-space summation of interaction contributions. Those are typically removed in the crudest manner by equating diverging terms to zero, causing finite-size errors that vanish slowly with the size of the computational supercell. \cite{MakovPayne1995}

In order to cancel plane-wave errors, electrostatic corrections are necessary. For conventional DFT approximations, many successful real-space countercharge corrections have been proposed. \cite{Blochl1995, Schultz2000, Schultz2006, DaboKozinsky2008, HineFrensch2009, HamadaOtani2009} Those improve computational accuracy at the cost of some consented increase in computational cost. However, in typical OD-DFT calculations, real-space countercharge corrections become costly due to the fact that OD-DFT functionals require to compute a different interaction potential, and thus a different correction, for each electron state at variance with Kohn-Sham DFT that requires instead the solution of one unique Hamiltonian problem per self-consistent iteration of the electronic-structure calculation.

Auxiliary-function techniques \cite{GygiBaldereschi1986, WenzienCappellini1995, JarvisWhite1997, MartynaTuckerman1999, MinaryTuckerman2002, RozziVarsano2006, Ismail-Beigi2006, CarrierRohra2007, SpencerAlavi2008, BroqvistAlkauskas2009} are reciprocal-space countercharge corrections that offer the advantage of not imposing any increase in computational cost in plane-wave OD-DFT calculations. Nevertheless, there exists much freedom in the possible choices of auxiliary functions and different definitions result in various levels of accuracy in eliminating plane-wave errors. Recently, Broqvist, Alkauskas, and Pasquarello have made an important step toward removing arbitrariness in the selection of auxiliary functions by introducing a simplified constant-shift correction. \cite{BroqvistAlkauskas2009} This constant-shift correction allows to describe energy observables with improved accuracy. However, the method is not intended to rectify properties that depend on the precise profile of the effective OD-DFT potential, such as the spatial confinement and electrical response of electron orbitals. 

In this work, we determine exact point-charge reciprocal-space auxiliary functions for the OD-DFT description of molecular systems exhibiting arbitrary translational symmetries.

The study is organized as follows. First, we recall the methodological framework of reciprocal-space auxiliary-function corrections. Second, we examine auxiliary-function corrections for nonperiodic and two-dimensional molecular structures, thereby recovering existing corrections in simple explicit forms. We then address the untreated one-dimensional case. In the last section, we assess the precision of the exact point-charge auxiliary-function correction in describing the electronic levels and electrical response of semiconducting oligomers of finite and infinite lengths.

\section{Method}

Plane-wave electronic-structure calculations for materials that do not exhibit three-dimensional periodicity are typically carried out within the supercell approximation, \cite{PayneArias1992} which consists of calculating the physical properties of an electronic system in a periodic finite cell and extrapolating computed observables in the infinite-cell limit where interactions between the system and its artificial periodic images can be safely neglected. In this section, we present the framework of reciprocal-space auxiliary-function countercharge corrections for the cancellation of periodic-image errors in supercell calculations. 

In order to put matters into perspective, we consider a nonperiodic charge density $\rho_{0\rm d}({\bf r})$ confined within a supercell of volume $\Omega$. Its three-dimensional periodic counterpart is written $\rho_{3 \rm d}({\bf r})=\sum_{\bf R} \rho_{0\rm d}({\bf r}+{\bf R})$ where the vectors $\bf R$ describe invariant translations of the superlattice. The Fourier components of the charge density read
\begin{equation}
\rho_{3\rm d}({\bf g})=\frac{1}{\Omega} \int d{\bf r}  \rho_{0 \rm d}({\bf r}) e^{-\i \bf g r}=\frac{1}{\Omega} \int_{\rm C} d{\bf r}  \rho_{3\rm d}({\bf r}) e^{-\i \bf g r},
\label{RhoF}
\end{equation}
where C stands for the supercell. Within the supercell approximation, the potential $v_{3 \rm d}({\bf r})$ is obtained by solving algebraically the Poisson equation in reciprocal space, yielding $v_{3 \rm d}({\bf g})=\frac{4\pi}{g^2}\rho_{3 \rm d}({\bf g})$ with the notable exception of the component $v_{3 \rm d}({\bf g}={\bf 0})$  that is singular due to the diverging factor. This singularity is conventionally removed by simply omitting the singular component, i.e.,
\begin{equation}
v_{3 \rm d}({\bf r})=\sum_{{\bf g} \neq {\bf 0}} \frac{4\pi}{g^2} \rho_{3 \rm d}({\bf g})e^{\i\bf gr}.
\label{VPW}
\end{equation}
Physically, this crude correction can be interpreted as immersing the charge density in an artificial compensating jellium. \cite{MakovPayne1995}

In contrast to the scheme presented above, auxiliary-function countercharge corrections do not rely on adding compensating jellium contributions to the charge density. They consist instead of directly transforming the periodic electrostatic kernel $\frac{4\pi}{g^2}$ by inserting an auxiliary function $\phi({\bf g})$ into the reciprocal-space sum:
\begin{equation}
v_{3\rm d}({\bf r})=\sum_{{\bf g}} \left(\frac{4\pi}{g^2} + \Delta \phi({\bf g})\right)\rho_{3 \rm d}({\bf g})e^{\i\bf gr}.
\label{VAF}
\end{equation}
In Eq.~(\ref{VAF}), the correction $\Delta \phi({\bf g})$ is defined in terms of the auxiliary function $\phi({\bf g})$ as
\begin{equation}
\Delta \phi({\bf g})=\frac{1}{\Omega} \int_{\rm WS} d{\bf r}  \phi({\bf r}) e^{-\i \bf g r} - \frac{4\pi}{g^2} \varrho({\bf g}),
\label{AFC2}
\end{equation}
where  $\varrho({\bf g})$ represents the periodic Fourier decomposition of the auxiliary density $\varrho({\bf r})=-\frac{1}{4\pi} \nabla^2 \phi({\bf r})$ and WS stands for the Wigner-Seitz cell centered around the origin. 

The auxiliary function must be chosen in order to cancel divergence singularities, and in such a manner that $v_{3\rm d}({\bf r})$ closely reproduce the potential of the original nonperiodic density
\begin{equation}
v_{0 \rm d}({\bf r})=\int d{\bf r}' \frac{\rho_{0 \rm d}({\bf r'})}{|{\bf r}-{\bf r}'|}.
\end{equation} 
The appropriate auxiliary function  can be determined by considering a point charge $\rho_{0 \rm d}({\bf r})=\delta({\bf r})$. In this simple case, it is straightforward to show that the choice
\begin{equation}
\phi({\bf r})=\frac \Omega r
\label{EAF}
\end{equation}
yields the exact potential everywhere within the Wigner-Seitz cell. The resulting simple plane-wave correction can be expressed as
\begin{equation}
\Delta \phi({\bf g})=\int_{\rm WS} d{\bf r}  \frac{e^{-\i \bf g r}}r - \frac{4\pi}{g^2},
\label{PCC2}
\end{equation}
which effectively removes the divergence singularity. 

Then, making use of the principle of superposition, it can be shown that the corrected potential $v_{3 \rm d}({\bf r})$ is also exact for any given charge density $\rho_{0 \rm d}({\bf r})$ under the requirement that the diameter of the arbitrary density $\rho_{0 \rm d}({\bf r})$ does not exceed half of that of the Wigner-Seitz cell, an important condition that is always tacitly assumed in the framework of reciprocal-space corrections.

Now, the central difficulty in making use of Eq.~(\ref{PCC2}) is evaluating the integral over WS due to the nonspherical WS geometry and the divergence of the integrand at the origin. In order to allow the evaluation of the reciprocal-space correction, one alternative  ansatz consists of restricting the exact point-charge potential to a sphere of radius $r_{\rm c}$ centered at the origin: \cite{JarvisWhite1997, RozziVarsano2006, Ismail-Beigi2006}
\begin{equation}
\phi({\bf r})=\left\{
\begin{array}{l} 
 \frac \Omega r ~ {\rm if } ~ r\le r_{\rm c}\\ 
0 ~ {\rm if} ~ r>r_{\rm c}.
\end{array}
\right.
\label{EAR}
\end{equation}
By Fourier transform, it can be shown that Eq.~(\ref{EAR}) amounts to choosing
\begin{equation}
\Delta \phi({\bf g})=-\frac{4\pi}{g^2} \cos(gr_{\rm c}),
\label{AFCT}
\end{equation}
which also cancels out divergence singularities. \footnote{Strictly speaking, truncation corrections do not belong to the family of auxiliary-function corrections as they cannot be cast into the form of Eq.~(\ref{AFC2}). Nevertheless, they can still be written as reciprocal-space corrections [Eq.~(\ref{VAF})].} However, Eq.~(\ref{AFCT}) presents the notable disadvantage of contributing oscillatory cosine terms to the potential $v_{3 \rm d}({\bf r})$ as a result of the sharp truncation at $r=r_{\rm c}$. Such oscillatory contributions affect the convergence of calculated electronic observables as a function of the size of the supercell and plane-wave kinetic-energy cutoff. Additionally, it should be noted that truncation schemes cannot be efficiently applied to elongated structures (see Sec. \ref{ApplicationSection}).

Therefore, in order to remove singularities without affecting numerical convergence, another widely used regularization technique consists of applying a Gaussian smearing to Eq.~(\ref{EAF}). \cite{MartynaTuckerman1999} Explicitly, the resulting Gaussian auxiliary function reads
\begin{equation}
\phi_\sigma({\bf r})=\frac \Omega r {\rm erf}\left( \frac r \sigma \right),
\label{GAF}
\end{equation}
that is,
\begin{equation}
\Delta \phi_\sigma({\bf g})=\int_{\rm WS} \frac {d{\bf r}}r {\rm erf}\left( \frac r \sigma \right){e^{-\i \bf g r}} - \frac{4\pi}{g^2} e^{- \frac{g^2 \sigma^2}{4}}.
\label{GCC2}
\end{equation}
The Gaussian correction defined in Eq.~(\ref{GCC2}) is well behaved, correctly removes singularities in plane-wave sums, and does not introduce oscillatory components in the computation of $v_{3 \rm d}({\bf r})$. Moreover, its computational precision can be improved systematically by decreasing the spread parameter $\sigma$.

Nevertheless, by examining the analytical behavior of Gaussian auxiliary-function corrections as a function of $\sigma$, Broqvist, Alkauskas, and Pasquarello have evidenced that the rate of convergence of the correction in the vanishing spread limit is poor (see Fig. 1 in Ref.~\onlinecite{BroqvistAlkauskas2009}). Moreover, computing the limit is costly due to the very fine grids required in representing Gaussians of vanishing spread, so that, in practice, the evaluation of the exact correction can only be performed for a restricted number of interpolation points. In fact, in the constant-shift correction of Broqvist, Alkauskas, and Pasquarello, the evaluation of the limit is only done at ${\bf r}={\bf 0}$, i.e.,
\begin{equation}
\Delta \phi({\bf g})
=\left\{
\begin{array}{l} 
\displaystyle\lim_{\sigma \to 0} \left( \displaystyle\frac{2\Omega }{\sqrt \pi \sigma} - \sum_{{\bf g}\neq {\bf 0}} \frac{4\pi}{g^2} e^{- \frac{g^2 \sigma^2}{4}}\right) ~ {\rm if } ~ g=0 \\ 
0 ~ {\rm if} ~ g \neq 0.
\end{array}
\label{CSC}
\right.
\end{equation}
This simple constant-shift approach has been shown to improve the convergence of hybrid-functional total energies and orbital levels as a function of supercell parameters. However, as already mentioned, the constant-shift auxiliary-function correction is not apt at and admittedly not intended to correcting electronic properties that depend on the precise profile of interaction potentials, such as the confinement of electronic orbitals and their response to an electric-field perturbation. In Sec.~\ref{PointChargeSection}, we derive point-charge auxiliary-function corrections for the precise determination of electrostatic and exchange interaction contributions.

\section{Point-charge auxiliary functions}

\label{PointChargeSection}

In this section, we present our computational procedure to obtain exact point-charge auxiliary functions for nonperiodic and periodic structures.  The derivation consists of examining the convergence of Gaussian auxiliary-function corrections [Eq.~(\ref{GCC2})] in the vanishing spread limit where the correction becomes exact. \footnote{It is important to note that the reciprocal-space point-charge and Gaussian corrections discussed here are  not related to the real-space point-charge and Gaussian countercharge corrections presented in Ref.~\onlinecite{DaboKozinsky2008}.}

This analysis allows to compute point-charge auxiliary functions at minimal computational cost. We first apply this simplified approach to recover the nonperiodic and two-dimensional corrections of Refs.~\onlinecite{MartynaTuckerman1999} and \onlinecite{MinaryTuckerman2002}. We then address the one-dimensional case, thereby complementing the derivation of point-charge auxiliary-function corrections for systems of arbitrary periodicity.

\subsection{Nonperiodic and two-dimensional molecular structures}

\begin{figure}[ht!]
\includegraphics[width=8.5cm]{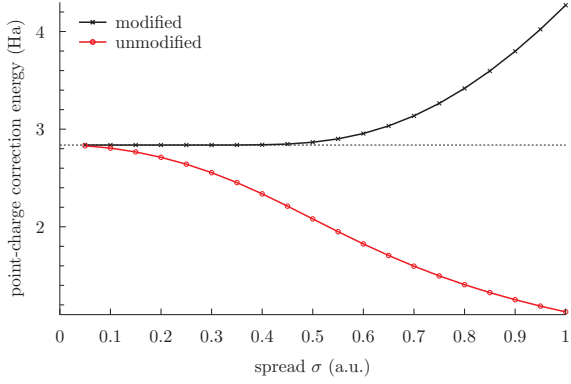}
\caption{Energy correction for a point charge in a cubic cell of volume $\Omega=1$ a.u. with and without point-charge modification term as a function of the spread parameter $\sigma$.  The auxiliary-function correction with point-charge modification converges rapidly to the exact Madelung energy of a point charge immersed in a jellium $\frac{\alpha_0}\Omega=2.837297$ Ha.
\label{ConstantConvergence}}
\end{figure}

We first focus on the asymptotic behavior of $\Delta \phi_\sigma({\bf r})$, which can be determined by differentiating the Fourier transform of Eq.~(\ref{GCC2}) with respect to the squared spread, obtaining
\begin{equation}
\frac{\partial \Delta \phi_\sigma({\bf r})}{\partial (\sigma^2)}=-\pi+\frac{\Omega}{\sqrt{\pi} \sigma^3} \sum_{{\bf R} \neq {\bf 0}} e^{-\frac{\left({\bf r}+{\bf R} \right)^2}{\sigma^2}}.
\label{AFCD}
\end{equation}
In Eq.~(\ref{AFCD}), one can observe that the sum term vanishes exponentially as $\sigma$ goes to zero. This allows to write a precise modified approximation for exact point-charge corrections in terms of their Gaussian counterparts:
\begin{equation}
\Delta \phi_{\sigma=0}({\bf r})=\Delta \phi_\sigma({\bf r})+\pi \sigma^2 + \cdots.
\label{AFCC}
\end{equation}
Equation~(\ref{AFCC}) corresponds to a simple modification of the original Gaussian auxiliary function [Eq.~(\ref{GAF})]:
\begin{equation}
\phi_\sigma({\bf r})=\pi \sigma^2 + \frac \Omega r {\rm erf}\left( \frac r \sigma \right).
\label{MGAF}
\end{equation}
The simplified derivation presented above underscores the significance of the quadratic term $\pi \sigma^2$ that enters into the expression of the original Martyna-Tuckerman correction. This constant contribution, which we refer to as the point-charge modification term, is only briefly alluded to in Ref.~\onlinecite{MartynaTuckerman1999} and is frequently omitted in practical implementations of auxiliary-function schemes.\cite{BroqvistAlkauskas2009} To illustrate the importance of the point-charge modification term, we compare the convergence of the modified auxiliary function [Eq.~(\ref{MGAF})] to that of the unmodified Gaussian auxiliary function [Eq.~(\ref{GAF})] in Fig.~\ref{ConstantConvergence}, demonstrating the considerably improved convergence of the modified correction to the exact point-charge limit for a test charge $\rho_{0 \rm d}({\bf r})=\delta({\bf r})$ in a cubic cell of unit volume.

We now turn to the two-dimensional case. Similarly to Eq.~(\ref{MGAF}), the expression of the two-dimensional correction reads
\begin{equation}
\phi_\sigma({\bf r})=\pi \sigma^2 + \Omega \varphi_{2 {\rm d}, \sigma}({\bf r}),
\label{MGAF2D}
\end{equation}
where  $\varphi_{2 {\rm d}, \sigma}({\bf r})$ denotes the potential of a two-dimensional array of Gaussian charges that can be expressed by longitudinal Fourier transform as
\begin{equation}
\varphi_{2{\rm d},\sigma}({\bf r}) = \sum_{\bf g_\parallel} \varphi_{2{\rm d},\sigma}({\bf r_\perp};{\bf g_\parallel}) e^{\i {\bf g}_\parallel {\bf r}_\parallel}.
\label{FDA2}
\end{equation}
In Eq.~(\ref{FDA2}), the vectors ${\bf r}_\parallel$ and ${\bf r}_\perp$ stand for the longitudinal and transverse components of the position vector ${\bf r}={\bf r}_\parallel+{\bf r}_\perp$.
The Fourier components $\varphi_{2{\rm d},\sigma}({\bf r_\perp};{\bf g_\parallel})$ satisfy the equation
\begin{equation}
\left( \frac{\partial^2}{\partial r_\perp^2}  - g^2_\parallel \right) \varphi_{2{\rm d},\sigma}({\bf r_\perp};{\bf g_\parallel})=- \frac{4 \pi}S g_{1{\rm d},\sigma}({\bf r_\perp}) {e^{-\frac{g_\parallel^2 \sigma^2}{4}}},
\label{E2D}
\end{equation}
where $g_{1{\rm d},\sigma}({\bf r_\perp})$  denotes a one-dimensional Gaussian distribution of spread $\sigma$ and $S$ stands for the longitudinal surface area of the supercell. Now, one can determine the physical solution $\varphi_{2{\rm d},\sigma}({\bf r_\perp};{\bf g}_\parallel={\bf 0})$ of Eq.~(\ref{E2D}) to be
\begin{equation}
\varphi_{2{\rm d},\sigma}({\bf r_\perp};{\bf g_\parallel}={\bf 0})=-\frac{2\pi}S 
\left( r_\perp {\rm erf}\left( \frac {r_\perp}\sigma \right)+\frac{\sigma}{\sqrt \pi}e^{-\frac{r_\perp^2}{\sigma^2}}\right).
\label{TPG0}
\end{equation}
Additionally, in the general case where $g_\parallel>0$, the solution of Eq.~(\ref{E2D}) is also obtained straightforwardly through a Gaussian convolution involving the point-charge solution
\begin{equation}
\varphi_{2{\rm d},\sigma=0}({\bf r_\perp};{\bf g_\parallel})=-\frac{2\pi}S \frac{e^{-g_\parallel r_\perp}}{g_\parallel}.
\label{PS0}
\end{equation}
In explicit terms, the component  $\varphi_{2{\rm d},\sigma}({\bf r_\perp};{\bf g_\parallel})$ can be written as
\begin{eqnarray}
\varphi_{2{\rm d},\sigma}({\bf r_\perp};{\bf g_\parallel})&=& \frac{\pi}{Sg_\parallel}\left(2 {\rm cosh}(g_\parallel r_\perp) \right. \nonumber \\ &+&
e^{-g_\parallel r_\perp}{\rm erf}\left(\frac{r_\perp}{\sigma} - \frac {\sigma   g_\parallel} 2 \right)
\nonumber \\ &-& \left.
e^{g_\parallel r_\perp}{\rm erf}\left(\frac{r_\perp}{\sigma} + \frac{\sigma g_\parallel} 2 \right)
\right).
\label{PSC}
\end{eqnarray}
As a final result, the modified Gaussian auxiliary function for two-dimensional molecular structures reads
\begin{eqnarray}
\phi_\sigma({\bf r})& =&\pi \sigma^2 
-\frac{2\pi\Omega}S 
\left( r_\perp {\rm erf}\left( \frac {r_\perp}\sigma \right)+\frac{\sigma}{\sqrt \pi}e^{-\frac{r_\perp^2}{\sigma^2}}\right)
\nonumber \\ &+&\Omega \sum_{{\bf g}_\parallel \neq {\bf 0}} \varphi_{2{\rm d},\sigma}({\bf r_\perp};{\bf g_\parallel}) e^{\i {\bf g}_\parallel {\bf r}_\parallel}.
\label{AF2D}
\end{eqnarray}

\begin{figure}[ht!]
\includegraphics[width=8.5cm]{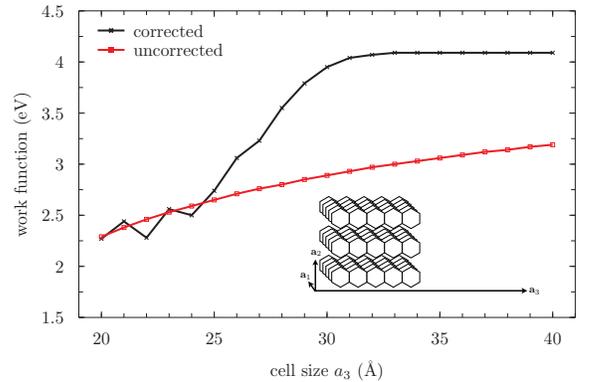}
\caption{PZ work function of a model pentacene film as a function of the longitudinal supercell parameter with and without point-charge auxiliary-function correction.
\label{PentaceneFilm}}
\end{figure}

To assess the performance of the two-dimensional point-charge auxiliary function, we calculate the work function of a model pentacene thin film consisting of one pentacene molecule in an orthorhombic supercell of fixed transverse dimensions $a_1=5$ \AA\ and $a_2=10$ \AA, and varying longitudinal dimension $a_3$. Here, we use  the Perdew-Zunger (PZ) OD-DFT self-interaction correction \cite{Perdew1981} and we employ norm-conserving local-spin density (LSD) pseudopotentials \footnote{The influence of substituting DFT pseudopotentials for their OD-DFT counterparts is discussed in Refs.~\onlinecite{BroqvistAlkauskas2009} and \onlinecite{DaboFerretti2010} and is expected to affect absolute electronic levels by less than 0.1-0.2 eV and absolute polarizabilities by at most a few tenths of an a.u., as supported by the comparison with refined GTO results in Sec~\ref{ApplicationSection}.} with a plane-wave energy cutoff of 40 Ry. (Further details on the application of point-charge auxiliary-function corrections to periodic systems are presented in Sec.~\ref{ApplicationSection}.) In Fig.~\ref{PentaceneFilm}, we observe that the convergence of the work function  without auxiliary-function correction is very poor. Applying the correction is found to improve convergence; expectedly, while corrected predictions are still unreliable for cell sizes comparable to thickness of the pentacene film, optimal accuracy (within a few meV) is achieved above 30 \AA, that is, approximately twice the thickness of the film, thereby demonstrating the performance of the two-dimensional auxiliary-function correction.

\subsection{One-dimensional molecular structures}

In this section, we examine molecular structures exhibiting one-dimensional translational symmetries. In the one-dimensional case, the modified Gaussian auxiliary function can be obtained from a simple generalization of Eq.~(\ref{MGAF}), i.e.,
\begin{equation}
\phi_\sigma({\bf r})=\pi \sigma^2 + \Omega \varphi_{1{\rm d},\sigma}({\bf r}),
\label{MGA1}
\end{equation}
where $\varphi_{1{\rm d},\sigma}({\bf r})$ is the electrostatic potential of a one-dimensional periodic array of Gaussian charges. As previously, the only potential difficulty in determining the point-charge auxiliary function  pertains to the computation of $\varphi_{1{\rm d},\sigma}({\bf r})$. The computational procedure is explained below.

The one-dimensional Gaussian potential  can be calculated from the Fourier ansatz
\begin{equation}
\varphi_{1 {\rm d},\sigma}({\bf r}) = \sum_{\bf g_\parallel} \varphi_{1 {\rm d},\sigma}({\bf r_\perp};{\bf g_\parallel}) e^{\i {\bf g}_\parallel {\bf r}_\parallel}.
\label{FDA}
\end{equation}
Indeed, substituting Eq.~(\ref{FDA}) into the Poisson electrostatic problem for a one-dimensional array of Gaussian charges, we obtain
\begin{eqnarray}
\left( \frac 1 {r_\perp} \frac{\partial}{\partial r_\perp} r_\perp \frac{\partial}{\partial r_\perp} - g^2_\parallel \right) \varphi_{1 {\rm d},\sigma}({\bf r_\perp};{\bf g_\parallel}) \nonumber \\ =- \frac{4 \pi}L g_{2 {\rm d},\sigma}({\bf r_\perp}) {e^{-\frac{g_\parallel^2 \sigma^2}{4}}},
\label{E2D}
\end{eqnarray}
which involves the transverse Gaussian function $g_{2 {\rm d},\sigma}({\bf r_\perp})=\frac{1}{\pi \sigma^2}e^{-\frac{r_\perp^2}{\sigma^2}}$ and the distance $L$ between periodic Gaussians. For the component corresponding to the vanishing longitudinal wavevector ${\bf g}_\parallel={\bf 0}$, the physically admissible solution reads \cite{DaboKozinsky2008}
\begin{equation}
\varphi_{1 {\rm d},\sigma}({\bf r_\perp};{\bf g_\parallel}={\bf 0})=\frac 1L \left(- \ln \left( \frac{r_\perp^2}{\sigma^2} \right) + {\rm Ei}\left( - \frac{r_\perp^2}{\sigma^2} \right)\right).
\label{PG0}
\end{equation}
In Eq.~(\ref{PG0}), Ei denotes the exponential-integral function  and  $\gamma$ stands for the Euler constant.

We can now focus on the more difficult case where  $g_\parallel>0$ for which it is convenient to introduce the rescaled radial coordinate $x=g_\parallel r_\perp$ and to study the solution  $k_\alpha(x)$ of the differential equation
\begin{equation}
\left( x^2 \frac{d^2}{dx^2} +  x \frac{d}{dx} - x^2 \right) k_{\alpha}(x)=- 2 \pi x^2 g_{2,\alpha}(x),
\label{K2D}
\end{equation}
which is related to the original solution $\varphi_{1{\rm d},\sigma}({\bf r}_\perp;{\bf g}_\parallel)$ through
\begin{equation}
\varphi_{1{\rm d},\sigma}({\bf r}_\perp;{\bf g}_\parallel)=2 k_{g_\parallel \sigma}(g_\parallel r_\perp) \frac{{e^{-\frac{g_\parallel^2 \sigma^2}{4}}}}{L}.
\end{equation}
The solution of Eq.~(\ref{K2D}) is slightly technical. Thus, for the sake of clarity, we defer the analytical derivation and numerical calculation of  $k_\alpha(x)$ to Appendix \ref{KD}.

Having determined the physical solution of Eq.~(\ref{K2D}), the final expression of the point-charge auxiliary function reads
\begin{eqnarray}
\phi_\sigma({\bf r}) &=&\pi \sigma^2 + \frac \Omega L \left(- \ln \left( \frac{r_\perp^2}{\sigma^2} \right) + {\rm Ei}\left( - \frac{r_\perp^2}{\sigma^2} \right)\right) \nonumber \\
&+&\frac {2\Omega} L \sum_{{\bf g}_\parallel \neq {\bf 0}} k_{g_\parallel \sigma}(g_\parallel r_\perp) {e^{-\frac{g_\parallel^2 \sigma^2}{4}}}e^{\i{\bf g}_\parallel {\bf r}_\parallel},
\label{AF1D}
\end{eqnarray}
which completes the explicit derivation of auxiliary functions for systems of arbitrary periodicity. 

The performance of the one-dimensional point-charge auxiliary function is discussed and demonstrated in Sec.~\ref{ApplicationSection}.

\section{Oligomers and polymers}

\label{ApplicationSection}

In this section, we evaluate the precision of point-charge auxiliary-function corrections in describing the electronic properties of periodic and nonperiodic molecular structures. 

At the implementation stage, we have found practical to collect point-charge auxiliary-function subroutines into a self-contained {\sc libafcc} (library of auxiliary-function countercharge corrections) module that takes as an input information about the geometry, periodicity, and grid resolution of the supercell and returns the correction calculated from Eqs.~(\ref{MGAF}), (\ref{AF2D}), and (\ref{AF1D}) on the real-space grid. The calculated correction is then added to the Coulomb kernel for the computation of electrostatic and exchange electron interactions, pseudopotential terms, and interatomic forces.

An early implementation of the point-charge auxiliary-function countercharge correction has been employed by the authors in Ref.~\onlinecite{DaboFerretti2010} for the OD-DFT description of isolated molecules and clusters. Here, we present a critical assessment of different auxiliary-function techniques in determining the electronic properties of extended molecular chains. Besides their known technological relevance to organic optoelectronics, \cite{VarsanoMarini2008} semiconducting polymers represent a critical test in assessing the predictive ability of electronic-structure methods. \cite{FaassenBoeij2002, KummelKronik2004, BaerNeuhauser2005, UmariWillamson2005, KorzdorferMundt2008, RuzsinszkyPerdew2008, VarsanoMarini2008} (In fact, conventional DFT approximations systematically destabilize donor charge-carrier levels in conjugated polymers and strongly overestimate their longitudinal electrical response and optical cross section.) \cite{ChandrossMazumdar1994, FaassenBoeij2002, VarsanoMarini2008}

We consider an isolated C$_6$H$_8$ molecule in a geometry recurrently studied in the literature, which consists of a conjugated carbon backbone with alternating simple and double bonds of 1.451 \AA\ and 1.339 \AA, respectively. \cite{KirtmanToto1995, FaassenBoeij2002, VarsanoMarini2008} 

We first focus on the supercell convergence of the ionization potential that is computed as the opposite energy of the highest occupied molecular level. In carrying out these first calculations, we employ norm-conserving LSD pseudopotentials with a plane-wave cutoff of 50 Ry for the Fourier expansion of the wave functions. Electronic optimization is achieved using Car-Parrinello fictitious damped dynamics, as implemented in the {\sc cp} code of the {\sc quantum-espresso} distribution. \cite{GiannozziBaroni2009} 

\begin{figure}[ht!]
\includegraphics[width=8.5cm]{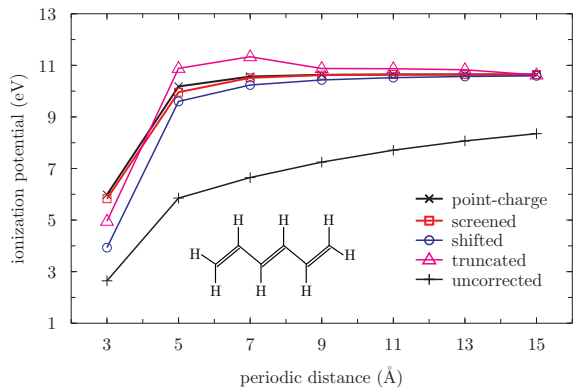}
\caption{PZ ionization potential of C$_6$H$_8$ as a function of the distance between periodic replicas with and without reciprocal-space corrections.
\label{C6H8IP}}
\end{figure}

Figure~\ref{C6H8IP} depicts the performance of auxiliary-function corrections in predicting the PZ ionization potential as a function of the periodic distance between artificial replicas. In Fig.~\ref{C6H8IP}, the truncated correction is that defined in Eq.~(\ref{AFCT}), the shifted correction is identical to the constant-shift correction of Eq.~(\ref{CSC}), and the screened correction is the auxiliary-function already available in the {\sc pw} code of {\sc quantum-espresso} (version 4.3.1), which we have adapted to the {\sc cp} code. Explicitly, this screened auxiliary-function correction reads
\begin{equation}
\Delta \phi({\bf g})
=\left\{
\begin{array}{l} 
2 \pi \sigma^2 ~ {\rm if } ~ g=0 \\ 
\Delta \phi_\sigma ({\bf g})  e^{-\frac{g^2\sigma^2}4} ~ {\rm if} ~ g \neq 0,
\end{array}
\right.
\end{equation}
where $\Delta \phi_\sigma ({\bf g})$ is defined in Eq.~(\ref{GCC2}) and the term $2 \pi \sigma^2$ can be obtained following the point-charge modification procedure described in Sec.~\ref{PointChargeSection}.

The results reported in Fig.~\ref{C6H8IP} confirm the slow convergence of the PZ ionization potential without correction. At a vacuum separation of 15 \AA, that is, three times larger than the length of C$_6$H$_8$, the predicted ionization potential is still underestimated by more than 2 eV. Applying the truncated correction is found to improve convergence substantially. However, as already mentioned above, the performance of the truncated method for extended systems is strongly limited by the fact that the spherical truncation diameter should not exceed the transverse size of the supercell, which is itself much smaller than the actual longitudinal extent of the oligomer.  The constant-shift correction results in a better cancellation of finite-size errors, effectively reducing the lack of convergence to a few tenths of an eV beyond a distance of 9 \AA. In the same range, the screened correction method is found to be remarkably precise with errors on the order of 0.02 meV relative to the estimated ionization energy. Optimal convergence is obtained with the point-charge correction [Eq.~(\ref{MGAF})] that halves finite-size errors in comparison to the screened correction above 9 \AA.

\begin{figure}[ht!]
\includegraphics[width=8.5cm]{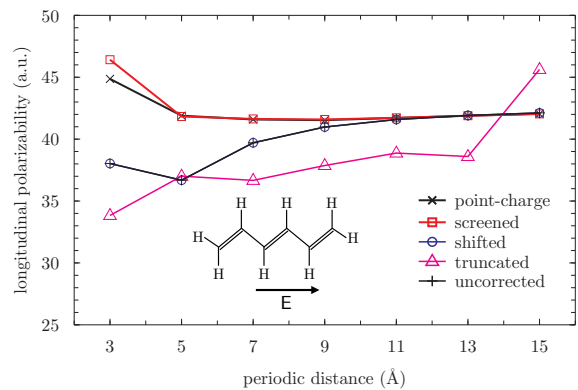}
\caption{PZ static longitudinal polarizability per monomer of C$_6$H$_8$ as a function of the distance between periodic replicas with and without reciprocal-space corrections.
\label{C6H8ElectricalResponse}}
\end{figure}

We now examine the static linear polarizability of C$_6$H$_8$ within PZ. In these calculations, the longitudinal polarizability is evaluated by finite difference from the static dipole moment computed at a field of 0.005 a.u. using Berry-phase techniques. \cite{King-SmithVanderbilt1993, Resta1992, Resta1994} The plane-wave cutoff energy and the oligomer geometry are kept unchanged. In Fig.~\ref{C6H8ElectricalResponse}, we observe that the truncated correction exhibits large numerical instabilities and that the constant-shift correction does not improve predicted polarizabilities relative to uncorrected calculations. The latter observation is explained by the fact that the constant shift does not affect the profile of the orbital-dependent potential. In contrast, both the point-charge and screened corrections improve electrical-response predictions. A closer comparison between the two methods turns to be in favor of the point-charge correction in the range 3-5 \AA\ and in favor of the screened correction in the range 9-15 \AA, thereby illustrating the potential benefit of screening the point-charge auxiliary-function correction to describe the electrical response of molecular systems. Nevertheless, as discussed above (Fig.~\ref{C6H8IP}), such improvement comes at the cost of altering the precision of predicted orbital energies so that the point-charge correction should be preferred in general.

\begin{table}
\caption{LSD and HF static longitudinal polarizability per monomer of oligoacetylenes as a function of the number of monomer with and without point-charge auxiliary-function correction, as compared with Slater-type-orbital (STO) and Gaussian-type-orbital (GTO) calculations.
\label{OligoacetyleneTable}}
\begin{ruledtabular}
\begin{tabular}{cccccccc}
\multicolumn{2}{c}{} & \multicolumn{6}{c}{$N$} \\
&&1&2&3&4&5&6 \\
\hline 
\multicolumn{2}{c}{Plane waves} \\
LSD & 60 Ry & 33.5 & 43.6 &	57.3 & 72.6 & 88.8 & 106.0 \\
HF  &	 60 Ry & 33.1 & 42.3 &	52.8 & 63.0 & 72.3 &  80.7 \\
\hline
\multicolumn{2}{c}{Local orbitals} \\						
LSD$^a$ & STO & 32 &	 42 & 56 & 71 & 87 & 105 \\
HF$^b$ & GTO 6-31G	& 37.4 & 47.3 & 57.2 & 66.4 & 74.7 & 82.2
\end{tabular}
\end{ruledtabular}
\flushleft
\footnotemark[1]{Reference \onlinecite{FaassenBoeij2002}.} \\
\footnotemark[2]{Reference \onlinecite{KirtmanToto1995}.}
\end{table}

To complement this comparison, we confront plane-wave corrected polarizability calculations to electrical predictions based upon local orbitals, namely, Slater-type orbitals (STOs) and Gaussian-type orbitals (GTOs). For the purpose of this analysis, it is necessary to raise the plane-wave energy cutoff to 60 Ry; with this computational parameter, we verify that the longitudinal polarizability of C$_6$H$_8$ is converged within less than one tenth of an atomic unit. Polarizabilities are again calculated with a Berry-phase electric field of 0.005 a.u. using the point-charge countercharge method. The results reported in Table~\ref{OligoacetyleneTable} illustrate the very good performance of the point-charge correction in reproducing Slater-type-orbital (STO) predictions. Nevertheless, it must be noted that there exist large deviations with respect to Gaussian-type-orbital (GTO) 6-31G Hartree-Fock predictions. A careful analysis of basis-set convergence reveals that the observed discrepancy is most likely due to the absence of polarization and diffuse functions in 6-31G. This interpretation is in line with the findings of Refs.~\onlinecite{Toto1995} and \onlinecite{ChampagneBotek2005} and is clearly corroborated by the calculations presented below.

In order to elucidate the origin of the discrepancy between plane-wave and local-orbital predictions, we now examine the static polarizability of finite dimerized hydrogen chains with alternating intramolecular and intermolecular distances of 2 a.u. and 3 a.u., respectively. The electrical response of dimerized hydrogen chains is the subject of a wide body of literature \cite{ChampagneMosley1995, KorzdorferMundt2008, FaassenBoeij2002, KummelKronik2004, BaerNeuhauser2005, UmariWillamson2005, RuzsinszkyPerdew2008, UmariMarzari2009} on assessing the relative performance of electronic-structure methods and describing optical saturation \cite{VarsanoMarini2008} in semiconducting polymers.

\begin{figure}[ht!]
\includegraphics[width=8.5cm]{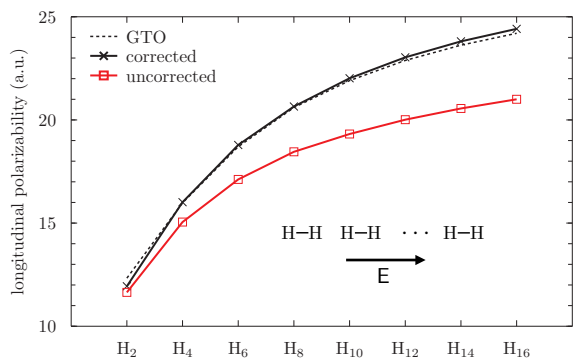}
\caption{HF static longitudinal polarizability per dimer of dimerized hydrogen chains with and without point-charge auxiliary-function correction, as compared with Gaussian-type-orbital (GTO) 6-311G(d,p) calculations.
\label{FiniteHydrogenChain}}
\end{figure}

Based upon GTO calculations, we can examine the absolute precision of our corrected plane-wave method in describing the electrical response of extended semiconducting oligomers. Our calculations employ the same parameters as those used in the calculations of Table~\ref{OligoacetyleneTable}. The vacuum separation between periodic chains is now of 60 atomic units. Figure~\ref{FiniteHydrogenChain} demonstrates the beneficial impact of the auxiliary-function method in predicting polarizabilities on a par with 6-311G(d,p) calculations within HF---and, by extension, within any type of hybrid functional admixture.

\begin{table*}
\caption{LSD, HF, and PZ static longitudinal polarizability per dimer of dimerized hydrogen chains as a function of the number of dimers with and without point-charge auxiliary-function correction, as compared with Gaussian-type-orbital (GTO) calculations at different levels of basis-set refinement. Periodic calculations are performed with 14 hydrogen units per supercell with a vacuum separation of $\sim$8 \AA. Extrapolated GTO results are given in parentheses.
\label{HydrogenTable}}
\begin{ruledtabular}
\begin{tabular}{ccccccccccc}
\multicolumn{2}{c}{} & \multicolumn{8}{c}{$N$} \\
&&1&2&3&4&5&6&7&8&$\infty$ \\
\hline 
\multicolumn{2}{c}{Plane waves} \\
LSD & 60 Ry & 12.4 & 18.7 & 24.1 & 28.5 & 32.1 & 35.2 & 37.7 & 39.9 & 55.4 \\
HF  & 60 Ry & 11.9 & 16.0 & 18.8 & 20.7 & 22.0 & 23.0 & 23.8 & 24.4 & 28.7 \\
PZ  & 60 Ry & 11.7 & 16.4 & 19.8 & 22.2 & 23.9 & 25.2 & 26.2 & 27.0 & 32.6 \\
\hline
\multicolumn{2}{c}{Gaussians} \\						
LSD & 6-311++G(d,p)$^a$ & --- & 18.8 & 24.3 & 28.8 & --- & 35.4 & 37.6 & --- & --- \\
HF & 6-31G(d,p)$^b$ & 11.3 & 15.3 & 18.1 & 20.1 & 21.5 & 22.5 & 23.3 & 23.9 & (28.5) \\
   & 6-311G(d,p)$^b$ & 12.3 & 16.0 & 18.7 & 20.6 & 21.9 & 22.9 & 23.6 & 24.2 & (28.6) \\
   & 6-311++G(d,p)$^a$ & --- & 16.1 & 18.9 & 20.8 & --- & 23.1 & 23.9 & --- & --- \\
PZ & 6-311++G(d,p)$^a$ & --- & 16.5 & 19.9 & 22.3 & ---	& 25.3 & 26.3 & --- & ---
\end{tabular}
\end{ruledtabular}
\flushleft
\footnotemark[1]{Reference \onlinecite{RuzsinszkyPerdew2008}.} \\
\footnotemark[2]{Reference \onlinecite{ChampagneMosley1995}.}
\end{table*}

To confirm this trend, we compare plane-wave LSD, HF, and PZ polarizabilities with GTO calculations in Table~\ref{HydrogenTable}. Here, the comparison of HF results is particularly enlightening; while 6-31G(d,p) underestimates the static longitudinal polarizability of dimerized hydrogen chains, predictions based upon 6-311G(d,p), which accounts for contributions from polarization functions, and 6-311G++(d,p), which incorporates the effect of additional diffuse functions, are found to be in very close agreement with corrected plane-wave calculations. The same level of agreement is obtained for LSD and PZ predictions with errors as low as 0.1-0.2 a.u. per dimer.

\begin{figure}[ht!]
\includegraphics[width=8.5cm]{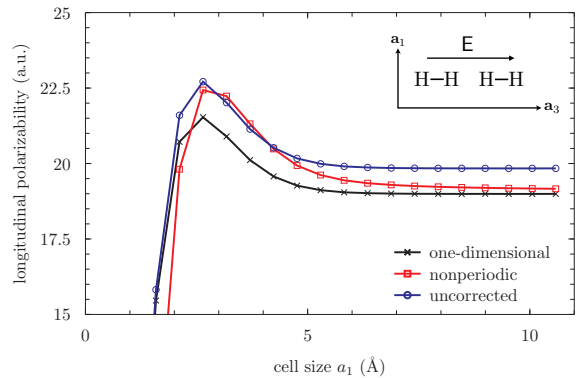}
\caption{HF static longitudinal polarizability per dimer of an infinite dimerized hydrogen chain (2 dimers per supercell) as a function of the distance between periodic replicas with nonperiodic and one-dimensional point-charge auxiliary function corrections.
\label{InfiniteHydrogenChain}}
\end{figure}

We conclude with the OD-DFT description of polarizability saturation in semiconducting polymers. \cite{VarsanoMarini2008} The converged OD-DFT description of the properties of infinite chains entails to treat nonperiodic orbital-dependent contributions to the effective potential separately from periodic orbital-independent contributions. In other words, the problem of determining the electronic states of periodic structures within OD-DFT methods is akin to describing a localized charge density confined in a neutral periodic host.

Explicitly, the electrostatic correction that we employ here (the one-dimensional auxiliary-function method) consists of computing contributions from the total periodic density using the one-dimensional correction [Eq.~(\ref{AF1D})] and contributions from individual orbitals using the nonperiodic correction [Eq.~(\ref{MGAF})]. This approach differs from the nonperiodic method that consists instead of computing periodic and nonperiodic contributions indiscriminately, using the same nonperiodic auxiliary function.

Figure~\ref{InfiniteHydrogenChain} compares the performance of the one-dimensional point-charge correction to that of the nonperiodic correction. Uncorrected results based upon the supercell approximation of Refs.~\onlinecite{UmariWillamson2005} and \onlinecite{RohraEngel2006}  are also reported. Here,  the HF electrical response of infinite hydrogen chains is determined keeping calculation parameters unchanged, with 2 hydrogen dimers per supercell to minimize computational cost. In Fig.~\ref{InfiniteHydrogenChain}, we observe that the nonperiodic and one-dimensional corrections exhibit similar convergence behaviors. Nevertheless, we also observe deviations of more than 2 a.u. between the two methods due to the inappropriate treatment of one-dimensional electrostatic and exchange contributions within the nonperiodic approach. 

The accuracy of one-dimensional calculations is further confirmed by direct comparison with extrapolated GTO calculations, as reported in the last column of Table~\ref{HydrogenTable}. As a matter of fact,  plane-wave Berry-phase predictions are found to be in very close agreement with extrapolated 6-311G(d,p) results\cite{ChampagneMosley1995} with deviations as low as 0.1 a.u., thereby establishing the very good precision of the point-charge auxiliary-function correction in describing extended oligomers and infinite polymers without extrapolation procedures.

\section{Conclusion}

In summary, we have presented a reciprocal-space computational method that relies solely on plane-wave techniques and allows to study nonperiodic and periodic molecular structures within the predictive framework of OD-DFT approximations. The approach employs reciprocal-space point-charge auxiliary-function corrections to achieve accurate convergence of electrostatic and exchange interactions at minimal computational cost. 

Because OD-DFT functionals require to compute a different interaction potential for each electron state at variance with DFT methods, any increase in the cost of evaluating interaction terms represents a potentially critical computational bottleneck, limiting the interest of elaborate real-space countercharge corrections.  

To derive point-charge auxiliary functions, we have examined the convergence of Gaussian auxiliary functions to the exact point-charge limit. In the process, we have highlighted the significance of corrective contributions that are frequently omitted in practical implementations. We have also demonstrated the performance of the method in describing the frontier levels and linear polarizability of molecular structures. Additionally, point-charge countercharge methods have been shown to put plane-wave OD-DFT calculations on a par with refined local-orbital calculations and to allow  the description of, e.g., infinite conjugated polymers without resorting to delicate extrapolation procedures.

In an effort to facilitate the incorporation of the point-charge auxiliary-function correction into conventional plane-wave codes, the correction has been implemented as a self-contained {\sc libafcc} module. \cite{Libafcc} In this form, we expect the method to prove useful in exploring further the accuracy of OD-DFT approximations for the description of periodic molecular structures.

\acknowledgments

The authors are indebted to Andrea Ferretti for valuable discussions. I.D. thanks Nicola Marzari, Stefano de Gironcoli, and Paolo Giannozzi for helpful comments and suggestions. 

The authors acknowledge support from Grant ANR 06-CIS6-014 of the French National Agency of Reseach. This work was undertaken as part of the postdoctoral appointment of  Y.L. at CERMICS, Universit\'e Paris-Est.

\appendix

\section{Derivation and computation of $k_\alpha(x)$} 

\label{KD}

In this appendix, we derive the analytical expression and explain the numerical calculation of the  solution $k_\alpha(x)$  of Eq.~(\ref{K2D}). The function $k_\alpha(x)$ is involved into the longitudinal Fourier expansion of the electrostatic potential  $\phi_{1 {\rm d}, \sigma}({\bf r})$ generated by a one-dimensional periodic array of Gaussian charges [Eq.~(\ref{FDA})].

Focusing first on the limit  $\alpha \to 0$ where the Gaussian function transforms into a Dirac distribution, the solution of Eq.~(\ref{K2D}) with vanishing asymptotic conditions reads 
\begin{equation}
k_0(x)=K_0(x),
\end{equation}
where $K_0(x)$ is the modified Bessel function of the second kind. Now, making use of the superposition principle, the solutions $k_{\alpha\neq 0}(x)$ can be obtained by Gaussian convolution:
\begin{equation}
k_\alpha(x)=\int d^{(2)}{\bf y} K_0(|{\bf x}-{\bf y}|) g_{2{\rm d},\alpha}(y).
\end{equation}
In practice, evaluating the above two-dimensional integral is difficult, except in the specific case where $x=0$. Making use of cylindrical coordinates, we can evaluate the function at $x=0$ to be
\begin{equation}
k_\alpha(0)=-\displaystyle\frac 12 e^{\frac{\alpha^2}4}{\rm Ei}\left(-\frac{\sigma^2}{4}\right).
\label{KS0}
\end{equation}
Now, the general solution of Eq.~(\ref{K2D}) can be computed from the conventional series expansion \cite{AbramowitzStegun1965}
\begin{equation}
k_{\alpha}(x)= \sum_{n=0}^{+\infty} \left( a_n + b_n \ln(x) \right) x^{2n}.
\label{KPS}
\end{equation}
Substituting Eq.~(\ref{KPS}) into Eq.~(\ref{K2D}) yields recursive relations of the form
\begin{equation}
\left\{
\begin{array}{cccc}
a_n & = & \displaystyle -\frac{b_n}{n} + \frac{a_{n-1}+c_{n-1}}{4n^2} \\
b_n & = & \displaystyle  \frac{b_{n-1}}{4n^2} & (n \ge 1),
\end{array}
\right.
\label{ABR}
\end{equation}
where the coefficients $c_n=\frac{2(-1)^n}{\alpha^{2n}(n-1)!}$ and $c_0=0$ are those entering into the expansion of the source term appearing on the right hand side of Eq.~(\ref{K2D}). Hence, the recursion defined by Eq.~(\ref{ABR}) allows to express  $k_\alpha(x)$ from the knowledge of the first terms of the sequence $a_0=k_\alpha(0)$ and  $b_0=0$ whose expression is obtained from Eq.~(\ref{KS0}).

\begin{figure}[ht!]
\includegraphics[width=8.5cm]{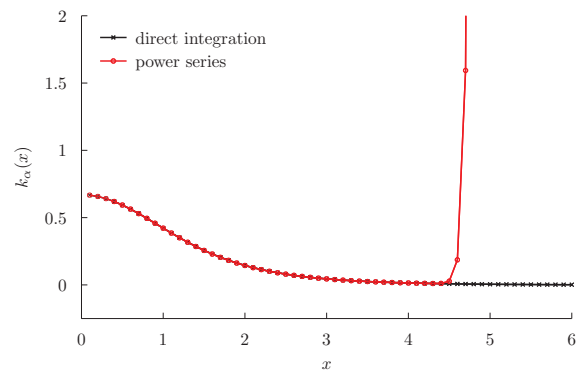}
\caption{$k_\alpha(x)$ computed by power series summation [Eq.~(\ref{KPS})] and direct Runge-Kutta integration [Eq.~(\ref{K2D})].
\label{GaussianK}}
\end{figure}

The dependence of $k_\alpha(x)$ on $x$ is depicted in Fig.~\ref{GaussianK}. We observe that the values computed from the series expansion [Eq.~(\ref{KPS})] vanish progressively upon increasing $x$ until reaching a critical point where double-precision computations become inaccurate. In this region of numerical inaccuracy, we instead resort to iterative integration techniques for the solution of Eq.~(\ref{K2D}) based upon the known asymptotic behavior
\begin{equation}
k_{\alpha}(x)=A_\alpha \sqrt {\frac{\pi}{2x}}e^{-x}+\cdots 
\label{AKA}
\end{equation}
of the function in the limit $x\to +\infty$. \cite{AbramowitzStegun1965} In Eq.~(\ref{AKA}), $A_\alpha$ is a constant that depends only on $\alpha$ and that can be determined straightforwardly from the requirement that $k_{\alpha}(x)$ should not diverge at the origin. This completes the computation of $k_{\alpha}(x)$.

\bibliography{article}

\end{document}